\documentclass[11pt,fleqn,twoside]{article}
\usepackage{amsfonts,amssymb,latexsym}
\makeatletter
\newcommand{\prava}[1]{\small\it
\begin{flushleft}
Copyright \copyright \ 1999 by  #1
\end{flushleft}}

\newcommand{\name}[1]{\begin{flushleft}
                       \LARGE \bf #1
                       \end{flushleft}\vspace{-3mm}}

\newcommand{\Author}[1]{\begin{flushleft}
                       \it #1 \end{flushleft}}

\newcommand{\Adress}[1]{\begin{flushleft}
                       \it #1 \end{flushleft}}

\newcommand{\Date}[1]{\begin{flushleft}
                      \small  \it #1 \end{flushleft}}

\newcommand{\ehkol}{Author \ name}
\newcommand{\ohkol}{Article \ name}
\renewcommand{\@evenhead}{
\hspace*{-3pt}\raisebox{-15pt}[\headheight][0pt]{\vbox{\hbox to \textwidth 
{\thepage \hfil \ehkol}\vskip4pt \hrule}}}
\renewcommand{\@oddhead}{
\hspace*{-3pt}\raisebox{-15pt}[\headheight][0pt]{\vbox{\hbox to \textwidth 
{\ohkol \hfil \thepage}\vskip4pt\hrule}}}
\renewcommand{\@evenfoot}{}
\renewcommand{\@oddfoot}{}

\newcommand{\be}{\begin{equation}}
\newcommand{\ee}{\end{equation}}
\newcommand{\ba}{\hspace*{-5pt}\begin{array}}
\newcommand{\ea}{\end{array}}

\newcommand{\ds}{\displaystyle}
\makeatother

\begin{document}

\thispagestyle{empty}
\setcounter{page}{273}
\renewcommand{\ehkol}{P. Guha and M. Olshanetsky}
\renewcommand{\ohkol}{Quest for Universal Integrable Models}

\begin{flushleft}
\footnotesize \sf
Journal of Nonlinear Mathematical Physics \qquad 1999, V.6, N~3,
\pageref{guha-fp}--\pageref{guha-lp}.
\hfill {\sc Article}
\end{flushleft}

\vspace{-5mm}

\renewcommand{\footnoterule}{} {\renewcommand{\thefootnote}{}
\footnote{\prava{P. Guha and M. Olshanetsky}}}

\name{Quest for Universal Integrable Models}\label{guha-fp}
\renewcommand{\thefootnote}{$^*$}
\Author{Partha GUHA~$^{\dag}$\footnote{
Present address: Mathematische Physik, Technische Universit\"at
Clausthal,
Arnold Sommerfeld Str. 6, D-38678 Clausthal-Zellerfeld, Germany
}
and Mikhail OLSHANETSKY~$^\ddag$}

\Adress{$\dag$~S.N. Bose National Centre for Basic Sciences, JD Block,\\
~~Sector -- 3,  Salt Lake,  Calcutta -- 700091, India;\\[1mm]
%\qquad \\[-2mm]
~~Institut des Hautes Etudes Scientif\/iques,  Le Bois-Marie,\\
~~35, Route de Chartres,  F-91440, Bures-sur-Yvette, France\\[1mm]
$\ddag$~Institute of Theoretical and Experimental Physics,
117259 Moscow, Russia}

\Date{Received September 02, 1998; Revised December 28, 1998;
Accepted January 08, 1999}

\begin{abstract}
\noindent
In this paper we discuss a universal integrable
model, given by a sum of two Wess-Zumino-Witten-Novikov (WZWN) actions,
corresponding to two dif\/ferent orbits of the coadjoint action of a
loop group on its dual, and the Polyakov-Weigmann cocycle 
describing their interactions.
This is an ef\/fective action for free fermions on a torus with
nontrivial boundary conditions.
 It is universal in the sense that all other known integrable
 models can be derived as reductions of this model.
Hence our motivation is to
present an unif\/ied description of dif\/ferent integrable models.
We present a proof of this universal action from the action of the
trivial dynamical system on the cotangent bundles of the loop group.
We also present some examples of reductions.
 \end{abstract}

\section{Introduction}
During the last two decades an essential progress has been achieved in
the investigation of integrable models [3, 5, 7, 8, 19].
Recently one of us [16] proposed an universal action for integrable
models. It turns out to be a sum of two Wess-Zumino-Witten-Novikov (WZWN)
 actions, corresponding to two
dif\/ferent orbits of the coadjoint action of a loop group,
and Polyakov-Weigmann cocycle~[20] describing their interaction.
The WZWN model is an universal object in the conformal
f\/ield theory. It is conjectured that all conformal
f\/ield theories are considered as some reductions of WZNW model
in the spirit of Drinfeld-Sokolov~[6], or of some
appropriate coset construction, and all the symmetries of the conformal
model are the symmetries of the WZNW model. In other words, all the algebraic
structures (operator algebras) arising in dif\/ferent conformal f\/ield
theories are considered as some reductions of the universal enveloping
Kac-Moody algebras. Hence the theory of $2d$ conformal models
is exhausted by the theory of the WZWN model.

Let $M$ be a closed two dimensional manifold and let $B$ denote a
three dimensional manifold with boundary $M$.

The WZWN action is given by
\[
 S_0 (g) = -\frac{k}{4\pi} \int_{M} d^2z \;
\mbox{Tr}\left(g^{-1}\partial g\right)^2
+ \frac{k}{12\pi} \int_B d^3y \epsilon^{ijk} \;
\mbox{Tr}\left(g^{-1}{\partial_i g}
g^{-1}{\partial_j g} g^{-1}{\partial_k g}\right),
\]
where $ g:M \longrightarrow {\hat G}$. 
The trace is the Killing form on the algebra
${\hat {\cal G}}$ of the loop group ${\hat G}$, and $k$ is the 
level of the af\/f\/ine algebra.

There are two dif\/ferent ways to derive the WZWN actions. Firstly,
it is an anamolous part of the ef\/fective action for fermions on a
plane in a gauge f\/ield~[20], and secondly it is obtained from the
Kostant-Kirillov form on an orbit of the coadjoint action of a loop
group~[3].
In this paper we shall search for a similar type of construction
for integrable models.
The action considered here is universal in the sense that all
known integrable models can be derived from it by reduction.
This action can be interpreted as an ef\/fective action for free
fermions on a torus with nontrivial boundary conditions,
where the role of perturbating relevant operators is played
by monodromies of the fermions.
  
This approach is useful from the point of view of string
theory -- the set of integrable models may play the role of
conf\/iguration space in string dynamics~[9, 13].

     In the late seventies M. Adler, B. Kostant and W. Symes [1, 4, 12, 24]
proposed a scheme
to construct integrable Hamiltonian systems. The AKS scheme originated
from their work and was subsequently developed by Reiman and
Semenov-Tian-Shansky [21, 22]. The scenario of the
 classical $R$-matrix was unveiled by
Semenov-Tian-Shansky~[22]. Recently one of us has proposed
a hierarchy of this formalism~[10].

Our approach is complimentary to the AKS formulation.
 The Euler-Lagrange
equation of motion of our proposed universal action, based on the Lie-Poisson
 structure, yields a zero
curvature equation. This is of course necessary, but does
 not fulf\/ill the suf\/f\/icient condition
of integrability. Only the choice of a special Hamiltonian, prescribed
by the Adler-Kostant-Symes scheme, guarantees the integrability in the 
Liouville sense.
Also, by choosing dif\/ferent coadjoint orbits and dif\/ferent matrix
entries, one can obtain various sets of integrable systems.    
Thus one can associate dif\/ferent integrable systems to various
symmetric spaces (see for example [11, 14, 17, 18]).

We organise this paper in the following way:
In section 2 we discuss some background material like the Hamiltonian action,
the moment map of the action of a loop group [23], etc.
We describe how some canonical dynamical
systems associate to a cotangent bundle of a Lie group [2].
A proof of our proposed 
universal action is presented in section three.
We derive this action from the action of the trivial dynamical
system on the cotangent bundles of the loop group.
In the f\/inal section we give some explicit examples.

Since there are no derivative terms appearing in the kinetic
part of the action, it seems that  we can not produce various
nontrivial mechanical systems. This is similar to the Hamiltonian system
on the cotangent bundle $(T^{\ast}G, \omega)$. The form $\omega$
is the symplectic form on the cotangent bundle.
This does not have an immediate mechanical
meaning (in the general sense). However, in both the cases they do
enable us to produce an interesting family of Hamiltonian systems
associated to a family of arbitrary Riemannian symmetric spaces. 
We explain this in one of our examples. 
We also add an appendix, where we present an explicit connection
between the nonlinear Schr\"odinger equation and the Heisenberg ferromagnetic
system, although this connection is known for some time (see for
example [17, 18]).

\section{Preliminaries}

\subsection{Hamiltonian action and moment map}

Let us start this section with some standard def\/initions
of Hamiltonian mechanics~[15].

Let $G$ be any compact semi-simple Lie group, ${\cal G}$ its
Lie algebra, and ${\cal G}^{\ast}$ the dual space of ${\cal G}$.
The left and right translation
\[
L_g : h \longmapsto gh, \qquad  R_g : h \longmapsto hg
\]
induces a map
\[
dL_{g}^{\ast}(\mbox{or } dR_{g}^{\ast}): T_{g}^{\ast}G
 \longrightarrow T_{e}^{\ast}G \cong {\cal G}^{\ast}.
\]
Thus if $(g,\kappa_g ) \in T^{\ast}G$, where $\kappa_g \in T_{g}^{\ast}G$
is the coordinate of the f\/ibre, then
\[
\ba{l}
( g, \kappa_g ) \stackrel{L}\longmapsto ( g,l_i), \qquad
l_i = dL_{g}^{\ast}\kappa_g,
\vspace{2mm}\\
(g, \kappa_g) \stackrel{R}\longmapsto (g,r_i), \qquad
r_i = -dR_{g}^{\ast}\kappa_g,
\ea
\]
where $r_i$ and $l_i$ are related by 
\[
r_i = -\mbox{Ad}^{\ast}g (l_i).
\]

Hence we can identify
\[
T^{\ast}G \stackrel{L,R}\longrightarrow G \times {\cal G}^{\ast}.
\]

 Let ${\hat {\cal G}} = C^{\infty}(S^1 ,{\cal G})$ be the loop
algebra and ${\hat G}$ the corresponding loop group.
By left or right trivialization, induced from $T^{\ast}G$, we can
identify
\[
T^{\ast}{\hat G} \simeq {\hat G} \times {\hat {\cal G}}^{\ast}.
\]

Let us consider a Hamiltonian action
\[
{\hat G} \times T^{\ast}{\hat G} \longrightarrow T^{\ast}{\hat G},
\]
such that 
\[
L_h (g, l_i) = (hg , l_i), \qquad 
R_h (g, l_i) = \left(gh , \mbox{Ad}^{\ast}h^{-1}(l_i)\right).
\]
The Hamiltonian actions are given by
\[
\mu_{X}^{L} (g , l_i) = - \langle X, \mbox{Ad}^{\ast}g (l_i) \rangle,
\qquad 
\mu_{X}^{R} (g , l_i) =  \langle X, l_i \rangle,
\]
where $ X \in {\hat {\cal G}}$, $ g \in T^{\ast}{\hat G}$
 and $ l_i \in {\hat {\cal G}}^{\ast}$.

Then the corresponding moment maps associated to the Hamiltonian
action are
\[
\mu_{L} (g,l_i) = \mbox{Ad}^{\ast} g (l_i),
\qquad 
\mu_R (g, l_i) = l_i.
\]

 Hence by symplectic reduction the reduced phase space is
naturally identif\/iable with the coadjoint orbit.

Let $ \alpha \in {\hat {\cal G}}^{\ast}$ be a constant element
then there exist a canonical one form 
\[
\theta := \langle \alpha , g^{-1}dg \rangle
\]
and the symplectic form 
\[
\Omega := d \langle \alpha , g^{-1}dg \rangle
\]
on ${\hat {\cal G}}^{\ast}$.

We def\/ine a geometrical action $ S = \int \theta$ on
$T^{\ast}{\hat G}$ as a functional of trajectories on 
$T^{\ast}{\hat G}$. The symmetries of this geometrical actions
are:
\[
\ba{l}
\alpha \rightarrow \alpha, \qquad g \rightarrow h_R g,
\vspace{2mm}\\
\alpha \rightarrow h_{L}^{-1}\alpha h_L, \qquad
g \rightarrow gh_L ,
\ea
\]
where $h_L$ and $h_R$ are constant elements of ${\hat G}$.    

\medskip

\noindent
{\bf Principle Bundle Construction:} If ${\hat {\cal O}_{\nu}}$ is the
coadjoint orbit in ${\hat {\cal G}}^{\ast}$ through the point $\nu \in 
{\hat {\cal G}}^{\ast}$, then there is a natural canonical imbedding
\[
i_{\nu}: {\hat {\cal O}}_{\nu} \longrightarrow {\hat {\cal G}}^{\ast} .
\]
The left translation is already indentif\/ied by
$ T^{\ast}{\hat G} \equiv {\hat G} \times {\hat {\cal G}}^{\ast}$. Then
the pullback map $i_{\nu}^{\ast}(T^{\ast}{\hat G})|_{\hat {\cal O}_{\nu}}$,
with restriction of $T^{\ast}{\hat G}$ on ${\hat {\cal O}}_{\nu}$,
is the principle
bundle over ${\hat {\cal O}}_{\nu}$.
 Let $ {\hat G} \times {\hat {\cal O}}_{\nu} \longrightarrow {\hat {\cal O}}_{\nu} $ is the %%@
trivial bundle over ${\hat {\cal O}}_{\nu}$. Let $ f : {\hat G} \times {\hat {\cal O}}_{\nu} %%@
\longrightarrow {\hat {\cal G}}$ be an equivariant
function. 
Let $\alpha_g \in i_{\nu}^{\ast}(T^{\ast}{\hat G})$ and $dR_{g}^{\ast}
 (\alpha_g ) \in T_{e}^{\ast}{\hat G} \equiv {\hat {\cal G}}^{\ast}$
then it is easy to prove

\medskip

\noindent
{\bf Proposition 2.1.}
{\it $f(\alpha_g) =  dR_{g}^{\ast}(\alpha_g)$ is the moment
map associated to the action
of ${\hat G}$ on $T^{\ast}{\hat G}$.}

\medskip

In the next section we describe a canonical integrable system
associated with $T^{\ast}G$, which can easily be lifted to 
$T^{\ast}{\hat G}$.

\subsection{A Universal Integrable System on
{\mathversion{bold} $T^{\ast}G$}}

In this section we present a brief description on
the construction of a universal integrable system on the
cotangent bundle of the Lie group~[2].

The moment map $ \mu : T^{\ast}{G} \simeq {G}
 \times {\cal G}^{\ast} \longrightarrow {\cal G}$,
 associated with the Hamiltonian
action $ {G} \times T^{\ast}{G} \rightarrow T^{\ast}{G}$,
is a Poisson map whenever 
${\cal G}^{\ast}$ is endowed with a natural Poisson structure.

\medskip

\noindent
{\bf Def\/inition 2.2.} {\it
A bivector $\Lambda \in \wedge^2 {\cal G}$ is called a Poisson bivector
if it commutes with itself
\[
[ \Lambda , \Lambda ] = 0.
\]

Two Poisson bivectors $\Lambda_1$, $\Lambda_2$ are called compatible
if they commute with one another
\[
 [ \Lambda_1 , \Lambda_2 ] = 0.
\]
This is equivalent to the fact that any linear combination
$l \Lambda_1 + m \Lambda_2$ is a Poisson bivector. This is called
pencil of Poisson bivectors.}

\medskip

A well known example is the rigid body system.
In this case the moment map is a Poisson map 
\[
 \mu : T^{\ast}SO(3) \longrightarrow so(3)^{\ast},
\]
with the linear Poisson structure 
\[
\Lambda_{so(3)^{\ast}} = \epsilon_{ijk}p_i \partial_j \otimes \partial_k.
\]

We consider dif\/ferential $1$-form $\eta$ on ${\cal G}^{\ast}$,
which is annihilated by the natural Poisson structure
$\Lambda_{{\cal G}^{\ast}}$ on ${\cal G}^{\ast}$
associated with the Lie bracket. Such a form is called
a Casimir form.

\medskip

\noindent
{\bf Def\/inition 2.3.}
{\it We define the vector field by $\Gamma_{\eta} = \Lambda (\mu^{\ast}(\eta))$,
and the dynamical system by
\[
\ba{l}
g^{-1}{\dot g} = \eta (g,\alpha) = \eta (\alpha),
\vspace{2mm}\\
{\dot \alpha} = 0,
\ea
\]
where $\Omega = d\langle \alpha, g^{-1}dg \rangle.$}

\medskip
%%%%%%%%%%%%%%%%%%% !!!!!!!!!!!!!!!!!!!!!
This system can be integrated by quadratures on each
 level set, obtained by f\/ixing $\alpha$'s in
${{\cal G}}^{\ast}$,
so that this particular dynamical system coincides with
a one parameter group of the action of ${G}$ on that
particular level set.

Consider, for example, the rigid rotator $ \eta = fdH$, where $H =
\sum\limits_{i}p_{i}^{2}/2$ is the Hamiltonian and $ f = f(p)$ is an
arbitrary function. If $\{ X_i \}$ is the basis of $so(3)$. Then
it is not dif\/f\/icult to~see that
\[
\Gamma_{\eta} = f(p)p_i {\hat X_i},
\]
where ${\hat X_i}$ are left invariant vector f\/ields on $SO(3)$.
Hence, the dynamical system is given~by
\[
\ba{l}
g^{-1}{\dot g} = f(p)p_i X_i,
\vspace{2mm}\\
{\dot p_i} = 0.
\ea
\]

In particular, if we restrict ourselves to the abelian Lie group
$R^n$, then $ \mu : T^{\ast}R^n \longrightarrow (R^n)^{\ast}$,
induced by the natural action of $R^n$ on itself by a
translation, is a Poisson map.
Let $\eta = \nu_k dI^k $ be a one form on $(R^n)^{\ast}$ in terms
of action-angle variables. After pulling it back to $T^{\ast}R^n$
we obtain a vector f\/ield $\Gamma_{\eta} = \Lambda (\mu^{\ast}(\eta))$,
where $\Lambda$ is the canonical Poisson structure in the cotangent bundle. 
Then the associated equations of motion on $T^{\ast}R^{n}$ or
$T^{\ast}{\bf T}^n$ is
\[
{\dot I}^k = 0, \qquad {\dot \phi}_k = \nu_k.
\]

We can recover this from another point of view.
Let the $Ad^{\ast}$-invariance
function $ H: T^{\ast}G \longrightarrow R$ satisf\/ies
\[
H = \frac{1}{2} || g^{-1}{\dot g} ||_{G}^{2}.
\]
This is a free particle Hamiltonian.
Now it is easy to see that if $(g,g^{-1}{\dot g}) \in T^{\ast}G$,
the equation assumes the form
\[
{\dot {(g^{-1}{\dot g})}} = 0.
\]

If we assume
\[
H(g,g^{-1}{\dot g}) = \frac{1}{2} || g^{-1}{\dot g} ||_{G}^{2}
- \langle \mbox{Ad}_{g}^{\ast}\alpha, \beta \rangle
\]
for $\alpha \in {\cal G}^{\ast}$, the equation becomes
\[
{\dot {(g^{-1}{\dot g})}} = [ \mbox{Ad}_{g^{-1}}^{\ast}(\beta), \alpha ].
\]

This equation is the nontrivial part of the canonical
system of equations for the free particle of the
Hamiltonian system $(T^{\ast}G, \Omega , H)$.

\section{Universal Integrable Model}

Recently [16] Olshanetsky proposed an action based on WZWN theory
which has the following form
\[
S = S^u (A) - H(A),
\]
where 
\[
S^u (A) = 2 \int \mbox{tr}(u {\bar \partial}gg^{-1}) d^2 z + kS_{WZWN}.
\]
Here $H(A)$ is a Hamiltonian and $A$ is a current, given by
\[
A = g^{-1}ug + g^{-1}\partial g.
\]
It def\/ines a point on the coadjoint orbit through a point $(u,k)$
in ${\hat {\cal G}}^{\ast}$.

Let us assume, for simplicity, that $k=1$.
The equation of motion, based on the Lie Poisson structure
\[
({\bar \partial} - \mbox{ad}_{\bar A}^{\ast})A = 0,
\]
is given by
\[
{\bar \partial}A = [ {\bar A}, A] + \partial {\bar A},
\]
where ${\bar A} = \mbox{grad}\; H$, the gradient of the Hamiltonian.
This is a zero curvature equation, which is a necessary but not
suf\/f\/icient condition of integrablity. Only the special Hamiltonians
guarantee this distinguish property. We must emphasize here that
${\partial}/{\partial {\bar z}}$ arises along with the ``time''
derivative $\frac{\partial}{\partial z}$,
 when the central extension of the classical algebra is considered.

Conformal models are distinguished by their holomorphicity
property: these are theories of massless scalars with
the equation of motion
\[
{\bar \partial}A = 0, \qquad \mbox{and} \qquad  A = \partial \phi.
\]
We have already seen that in the WZWN model the role 
of $A$ is played by the Kac-Moody currents.

The Adler-Kostant-Symes scheme can be used to choose
 a particular subset of the zero curvature equation which are
integrable in the Liouville sense.

Apparantly there is a drawback in this model, there appear no
$z$ derivative in the kinetic part of the action. 
This is the ref\/lection of the lack of central charge in the
$R$-algebra. This is exactly the necessary condition 
for the description of ordinary integrable systems.
But we shall show how to overcome this dif\/f\/iculty by constructing
the mechanical systems associated to the Riemannian symmetric
spaces via Fordy-Kulish decomposition. 
  
\subsection{AKS Scheme and Zero Curvature Equations}

Let ${\hat {\cal G}} = gl(n,{\bf C}) \times {\bf C}[\lambda, \lambda^{-1}]$
be the loop algebra of a semi-inf\/inite formal Laurent series in ${\lambda}$
with coef\/f\/icients in $gl(n,{\bf C})$. For example, an element
$X(\lambda) \in {\hat {\cal G}}$ can be expressed as a formal 
series in the form 
\[
 X(\lambda) = \sum_{i=-\infty}^{m} x_i \lambda^i \qquad 
\forall \; x_i \in gl(n,{\bf C}). 
\]
The Lie bracket, with $Y(\lambda) = \sum\limits_{j=-\infty}^{l} y_j \lambda^j$,
is given by
\[
 [ X(\lambda), Y(\lambda)] = 
\sum_{k=-\infty}^{m+l} \sum_{i+j=k} [x_i, y_j]\lambda^k. 
\]
   
We def\/ine a nondegenerate bilinear two form on ${\hat {\cal G}}$
\[
\langle  A(\lambda), B(\lambda) \rangle := \mbox{Res}_{\lambda = 0} (\lambda^{-1}
A(\lambda) B(\lambda)) = \mbox{tr} (A(\lambda)B(\lambda))_0. 
\]

There is a natural splitting in the loop algebra 
${\hat {\cal G}} = {\hat {\cal G}}^+ \oplus {\hat {\cal G}}^-$,
where ${\hat {\cal G}}^+$ denotes the subalgebra of ${\hat {\cal G}}$,
given by the polynomial in $\lambda$, and ${\hat {\cal G}}^-$ is the subalgebra
of strictly negative series.

The above decomposition of ${\hat {\cal G}}$ do not correspond
to a global decomposition of the loop group ${\hat G}$, but we have a dense
open subset 
\[
{\hat G}^- {\hat G}^+ \subset {\hat G},
\]
consisting of all loops $\phi$ that can be factorized in the form
\[
 \phi = \phi^- \phi^+ 
\]
with $\phi^- \in {\hat {\cal G}}^-$, $\phi^+ \in {\hat {\cal G}}^+$.
We refer to this subset of ${\hat G}$ as the {\it big cell}.

Let us consider the Grassmannian like homogeneous space
${\hat {\cal G}}/{\hat {\cal G}}^+$. The image in
 ${\hat {\cal G}}/{\hat {\cal G}}^+$ of the complement of the big cell
in ${\hat {\cal G}}$ is a divisor in ${\hat {\cal G}}/{\hat {\cal G}}^+$.
It therefore corresponds to a holomorphic line bundle
\[
{\cal L} \longrightarrow {\hat {\cal G}}/{\hat {\cal G}}^+.
\]
We denote by $LG$ the automorphism group of ${\cal L}$.
The pullback of ${\cal L}$ to $LG$ is canonically trivial.
Hence $LG$ turns out to be the central extension of ${\hat G}$
by $C^{\times}$:
\[
 1 \longrightarrow {\bf C^{\times}} \longrightarrow LG \longrightarrow 
{\hat G} \longrightarrow 1. 
\]

The loop algebra 
\[
L{\cal G} = {\hat {\cal G}} \oplus {\bf C}
\]
satisf\/ies the following commutation relation
\[
[ (A(\lambda),a),(B(\lambda),b) ] := ([A,B](\lambda),\omega (A,B)),
\]
where $\omega (A,B)$ is the Maurer-Cartan ${\bf C}$-valued two
cocycle
\[
 \omega (A,B) = \int_{0}^{2\pi}( A , \partial_{\bar z} B), 
\]
which satisf\/ies
\[
\omega (A,[B,C]) + \omega (B,[C,A]) + \omega (C,[A,B]) = 0.
\]

$L{\cal G}$ is called the central extension of ${\hat {\cal G}}$,
obtained through $\omega$. In this particular case $L{\cal G}$ is
also called a Kac-Moody algebra on $S^1$.
 
In general, the map
\[
{\kappa}: {\hat {\cal G}} \rightarrow L{\cal G} 
\]
is not a Lie algebra homomorphism, but only its restriction to 
$ {\hat {\cal G}}^+$ is a Lie algebra homomorphism,
since the central extension term vanishes identically.
The corresponding induced map
\[
{\kappa}: {\hat G}^+ \longrightarrow LG 
\]
yields a canonical holomorphic trivialization of the part of the
f\/ibration lying over ${\hat {\cal G}}$.

Let $\phi = \phi^- \phi^+$ be an element
of the big cell. Then ${\kappa}{\phi}$ satisf\/ies
\[
{\kappa}(\phi) = \kappa (\phi^-) \kappa (\phi^+),
\]
where ${\kappa}({\phi})$ is the dense open subset
of $L{\cal G}$ that lies over the big cell of ${\hat {\cal G}}$.

We also def\/ine the bilinear form on $L{\cal G}$
\[
\langle (A,a),(B,b) \rangle = ab + \int_{s^1} \mbox{tr}(AB).
\]

Let $R \in \mbox{End}\;{\cal G}$ be the linear operator on ${\cal G}$.
The Kostant-Kirillov-Souriau $R$-bracket is given by
\[
 [ X , Y ]_R = \frac{1}{2} ( [RX,Y] + [X,RY] )   \qquad \forall \; X,Y \in {\cal G}. 
\]
This satisf\/ies the Jacobi identity if $R$-satisf\/ies the modif\/ied
Yang-Baxter equation.

\medskip

\noindent
{\bf Def\/inition 3.1.} {\it
Let $({\hat {\cal G}},R)$ be a double loop algebra on which we define
two algebraic structures. Suppose also that $\omega$ is a 2-cocycle
on ${\hat {\cal G}}$.
Then
\[
 \omega_R (X,Y) = \omega (RX,Y) + \omega (X,RY)
\]
is a 2-cocycle on ${\hat {\cal G}_R}$.}

\medskip

The gradient $ {\nabla F} : {\cal G}^{\ast} \rightarrow {\cal G}$ is def\/ined
by 
\[
 \frac{d}{dt}F(U + tV)|_{t=0} = \langle V , {\nabla F}(U) \rangle. 
\]

\noindent
{\bf Lemma 3.2.} {\it Let $H$ be an
ad-invariant function on ${\hat {\cal G}}^{\ast}$.
Then the gradient of $H$ satisfies
\[
\mbox{\rm ad}^{\ast}(R{\nabla H}(\alpha),a)(\alpha , 1) = 
(\mbox{\rm ad}^{\ast}(R{\nabla H}(\alpha))(\alpha)
+ R{\nabla H}^{\prime} , 0 ).
\]}

\medskip

\noindent
Sketch of the {\bf Proof}:
By using the identity
\[
\langle \mbox{ad}_{R}^{\ast}(X,a)(\beta,c) , (Y,b) \rangle + 
\langle (\beta,c) , \mbox{ad}_R (X,a)(Y,b) \rangle = 0,
\]
we obtain our result. \hfill $\blacksquare$

\medskip

\noindent
{\bf Def\/inition 3.3.} {\it
There exists a natural Poisson structure on the space $C^{\infty}({\hat {\cal G}}^{\ast}, {\bf %%@
C})$
of smooth real valued functions on ${\hat {\cal G}}^{\ast}$
\[
\{ \xi , \chi \}(U,c) = \int_{S^1} \mbox{\rm Tr}
\left(c\frac{d({\nabla \xi)}}{d\bar z} +
 [ {\nabla \xi} , {\nabla \chi} ], U\right)d\bar z   \qquad  \forall \; \xi \chi \in
 C^{\infty}({\hat {\cal G}}^{\ast}).
\]}

We observe that the central parameter $c$ is f\/ixed under the
coadjoint action of the group. So $L{\cal G}^{\ast}$ stratif\/ies
into Poisson submanifolds, corresponding to dif\/ferent values of the
parameter.

The dif\/ferential equation  appears from the ad-invariant
 condition, which should be satisf\/ied by the
gradients of the local Hamiltonians 
\[
 ({\partial}_{\bar z} - \mbox{ad}^*\; \alpha){\nabla H} = 0. 
\]
It is known that the good substitutes for 
local Hamiltonians are Casimir functions.  
Hence we choose Hamiltonian $ H = \frac{1}{2}\mbox{tr}(\alpha^2 ). $

\medskip

\noindent
{\bf Theorem 3.4.} {\it 
Let $\alpha$ be the orbit. The Hamiltonian equations of motion
 on the ${\hat {\cal G}}^{\ast}$,
generated by the gradient of the Hamiltonian $H$ (the ad-invariant
function), have the form
\[
\frac{d{\alpha}}{dz} = \frac{Rd({\nabla H})}{d\bar z} + [ R({\nabla H}), \alpha ].
\]}
  
\subsection{ Derivation of the Action}
Let us derive the generic form of the action for IM from the action of the 
trivial dynamical system on the cotangent bundle $ T^{\ast}{\hat G}$

Let $(g,m;u,n) \in T^{\ast}LG $. Then by the left action of ${\hat G}$
follows
\[
 g \stackrel{L}\rightarrow gh, \qquad  m \rightarrow m, \qquad
n \rightarrow n, \qquad \forall \; h \in {\hat G}.
\]
Since $m$, $n$ are invariant under the action of group $h \in {\hat G}$,
the action foliates $T^{\ast}LG$ into hyperplanes. Let us conf\/ine to a
particular hyperplane $ m=0$, $n=1$.
We again consider the two form $\omega (g)$ on ${\hat G}$
\[
\omega (g) = \int dz \; \mbox{Tr} \langle dgg^{-1},\partial (dgg^{-1}) \rangle . 
\]

Let us replace $u$ by a new f\/ield
\[
 h = P \exp \int dz^{\prime} u(z^{\prime}).
\]
 
\noindent
{\bf Def\/inition 3.5.} {\it
A symplectic form on $T^{\ast}{\hat G}$ is given by
\[
\Omega = \omega (g) + \omega (h) + 2 \int dz \langle  u, (dgg^{-1})^2 
\rangle .
\]}

The corresponding one form $\beta$ satisf\/ies $ d\beta = \Omega$.
We def\/ine a Hamiltonian
\[
H(v) = 2 \int d^2 z \langle v, (hg)^{-1}\partial (hg) \rangle.
\]
Hence the action is given by
\[
\ba{l}
\ds S = \int dz^{\prime} ( \beta - H )
\vspace{3mm}\\
\ds \qquad \simeq S_{WZWN}(h) + S_{WZWN}(g) + 2 \int d^2z \langle u,
dgg^{-1} \rangle - H. 
\ea
\]
After the gauge f\/ixing condition we arrive at
\[
 S = S^u (A) - H(v,A),
\]
where $ H(v,A) = \int d^2 z \langle v, A\rangle. $

It was demonstrated by Polyakov and Wiegmann [20], that 
an ef\/fective action from the fermionic Lagrangian on a plane
\[
{\cal L} = {\bar \psi}^L (\partial - A^u )\psi^L +
{\bar \psi}^R ({\bar \partial} - A^{\bar v})\psi^R
+ \frac{1}{2\alpha_0} \langle  A^u , A^{\bar v} \rangle
\],
 gives rise to a sum of the WZWN action in a gauge invariant form.
In this case we have
\[
\log \frac{\det (\partial - A^u) \det({\bar \partial} - A^{\bar v})}
{\det ( \partial - u) \det({\bar \partial} - v)}
 = S - 2 \int {d^2 z} \langle  u , v \rangle.
\]

\noindent
{\bf Proposition 3.6.} {\it 
The equation of motion, corresponding to $ S = S^u (A) - H(v,A)$
for $ H (v,A) = \int d^2 z \langle v,A\rangle $, is 
\[
 {\bar \partial}A = [ v , A ] + \partial v.
\]}

Suppose $g$ is any arbitrary element of the loop group.
Then there exist a gauge transformation
\[
 u \longrightarrow g^{-1}ug + g^{-1}\partial g.
\]
 
Hence the matrices $ u,v $ satisfy the zero curvature 
equation
\[
 {\bar \partial}u = [ v,u] + \partial v.
\]

We assume that $u$ and $v$ depend on a spectral parameter $\lambda$
which lives on a rational curve ${\bf CP}^1$
\[
\ds  u = u_0 + \sum_{j=1}^{m_1} \frac{u_j}{\lambda - a_j},
\qquad 
\ds  v = v_0 + \sum_{k=1}^{m_2} \frac{v_k}{\lambda - b_k}
\],
such that $u$ and $v$ satisfy the zero curvature condition.

Since the zero curvature condition is preserved under the gauge transformation
$ u \longmapsto u^g = g^{-1}ug + g^{-1}\partial g = A $,
the corresponding linear equations
\[
 \partial \psi = A^u \psi, \qquad  {\bar \partial}\psi = A^{\bar v}\psi
\]
of the zero curvature equation  satisf\/y

\medskip
 
\noindent
{\bf Proposition 3.7.} {\it If $\partial \psi = -u \psi$, then $A$
also satisfies the same  equation for $g^{-1}\psi$.}

\medskip

\noindent
Sketch of the {\bf Proof:}
\[
 \partial (g^{-1}\psi) = -g^{-1}{\partial g}g^{-1}\psi + 
g^{-1}{\partial \psi}  = -A(g^{-1}\psi). 
\hspace{163pt} \blacksquare
\]

If we consider fermions with monodromies, then due to
the zero curvature condition the left and right function
can be identif\/ied by $ \psi^L = \psi^R = \psi. $

We may regard $\psi$ as a function on ${\bf R}$ with values in ${\hat G}$.
Its value  at $ x = 2\pi$ is called the monodromy matrix $T_A$.
The coadjoint orbits are described by Floquet's theorem.

\medskip

\noindent
{\bf Theorem 3.8.} (Floquet)
{\it Two periodic potentials $A$ and $A^{\prime}$ are gauge equivalent if and only
if the corresponding monodromy matrices $T_{A}$, $T_{A^\prime}$ are conjugate.}

\medskip

Consider some particular cases:

\medskip

\noindent
{\bf Remark 3.9.} {\it For the generic integrable models, the
following relations hold
automatically 
\[
 {\bar \partial}u = \partial v = [ u, v] = 0.
\]}

\noindent
{\bf Proposition 3.10.} {\it If an integrable model satisfies ${\bar \partial}u = \partial v =
 [ u , v] = 0$, then the zero curvature equation reduces to
\[
{\bar \partial}g = gv - vg.
\]}

\noindent
{\bf Proof.} We know that
\[
\ba{l}
A = g^{-1}ug + g^{-1}\partial g,
\vspace{2mm}\\
{\bar \partial} A = -g^{-1}{\bar \partial}gA +
 g^{-1}u{\bar \partial}g + g^{-1}\partial{\bar \partial}g .
\ea
\]
Let us substitute our ansatz $ {\bar \partial}g = gv - vg $
in the above equation:
\[
\ba{l}
 {\bar \partial}A = -g^{-1}{\bar \partial}g A + g^{-1}ugv - g^{-1}uvg
+ g^{-1}{\partial g}v - g^{-1}v{\partial g} 
\vspace{2mm}\\
\qquad  = -g^{-1}{\bar \partial}gA + Av - g^{-1}uvg - g^{-1}v{\partial g}
\vspace{2mm}\\
\qquad  = -g^{-1}(gv - vg)A + Av - g^{-1}uvg - g^{-1}v{\partial g}.
\ea
\]
Since 
\[
 g^{-1}vgA = g^{-1}vug + g^{-1}v{\partial g},
\]
we get back the equation of motion ${\bar \partial}A = [A,v].$
\hfill $\blacksquare$

\medskip

Additionally we have 
\[
 \partial g = gA - ug,
\]
which follows from the current.

\medskip

\noindent
{\bf Proposition 3.11.}
\[
 [ \partial , {\bar \partial} ]g = 0, \qquad \partial^2 \neq 0 \neq
{\bar \partial}^2. 
\]

\noindent
Sketch of the {\bf Proof:} Since
\[
{\bar \partial}g = gv - vg
\],
it is easy to see that
$ \partial {\bar \partial}g = {\bar \partial}{\partial}g.$
\hfill $\blacksquare$ 

\medskip

Let us consider the factorization of the matrix $g(\lambda)$ such
that 
\[
g(\lambda ) = g_{+}(\lambda)g_{-}^{-1}(\lambda)
\]
is the solution to the Riemann problem, where $g_+$ is an element of
 the group
of all smooth functions from the unit circle $S^1$ to $G$ that
extend to holomorphic $G$-valued functions on the disk
 $\{ \lambda : |\lambda| < 1 \}$.
Similarly, 
 $g_{-}^{-1}(\lambda)$ is the element of the 
group of all smooth functions
$ S^1 \rightarrow G$ that extend holomorphically to the disk
$\{ \lambda : |\lambda| > 1 \}$ and take the value $1$ at inf\/inity.

Substituting $ g = g_+ g_{-}^{-1}$ in $ {\bar \partial}g = gv - vg$ 
 and ${\partial g} = gA - ug$. We obtain
\[
\ba{l}
 g_{-}^{-1}vg_{-} + g_{-}^{-1}{\bar \partial}g_- 
= g_{+}^{-1}vg_+ + g_{+}^{-1}{\bar \partial}g_+,
\vspace{2mm}\\
g_{-}^{-1}Ag_- + g_{-}^{-1}{\partial g_-} =
g_{+}^{-1}ug_+ + g_{+}^{-1}{\partial g_+}.
\ea
\]

\noindent
{\bf Def\/inition 3.12.} {\it We define two new currents
\[
\ba{l}
 A^u = g_{+}^{-1}ug_+ + g_{+}^{-1}{\bar \partial}g_-,
\vspace{2mm}\\
 A^{\bar v} = g_{-}^{-1}vg_- + g_{-}^{-1}{\bar \partial}g_-.
\ea
\]}

Consider a contour $\gamma$ which consists of small
circles around the points $a_j$ $\!(j=1,\ldots, m_1)$.
Let us modify the action $S$ by
\[
 S \longrightarrow \int_{\gamma} d{\lambda} S. 
\]
Originally, Olshanetsky [???]  generalized $S$ by 
introducing the kinetic term ${\hbar }{\partial}_{\lambda}$ in  
such a way that one obtains, in addition to the zero curvature equation,
a new equation
of motion in the form of the string equation:
\[
 [ \partial + A(g) , {\hbar }\partial_{\lambda} + v^{\prime} ] = {\hbar }.
\]
Earlier, Gerasimov {\it et al} [8, 9]  proposed a number of 
of programs for incorporating integrable models into the general framework 
of string theory. The string theory is understood as some dynamical
theory on some conf\/iguration space which contains at least
all the 2-dimensional f\/ield theories as its points.
They argued that for the universal description of all conformal
models, it is necessary to treat various Kac-Koody algebras on the
same ground, through their embedding into ${\hat GL}(\infty)$
algebra. It may be described through dependence on some auxiliary
variable $\lambda$. Hence this explains why $\lambda$ appears in
the equation.

A string equation is a sort of ``quantum deformation'' of a zero curvature
equation. Uptil now the holomorphic dependence of the spectral parameter
is quite artif\/ical. Moreover, the geometrical meaning of this ``deformed''
zero curvature equation is still lacking. We now try to give a plausible
explanation of this equation.
 
So far we have encountered three coordinates $(z, {\bar z}, \lambda)$,
where ${\bar z}$ appears along with $z$ and
${\bar z}$ plays the role of ``time''. Apparantly it seems that
the lack of ${\bar \lambda}$ dependence may be a drawback
from the point of view of integrable systems. But this can be
 managed in the following way:

 Let $(z, \lambda, {\bar z}, {\bar \lambda})$ be the coordinates
on ${\bf R}^4$, which are independent and real for signature~(2,2).
The self dual Yang-Mills equations are the compatibility
conditions for the pair of operators
\[
L_0 = (D_z - \xi D_{\bar \lambda}), \qquad
  L_1 = (D_{\lambda} + \xi D_{\bar z}),
\]
where $\xi \in {\bf C}$ is an auxiliary complex 
spectral parameter and $D_z$ is the covariant derivative
of some Yang-Mills connection in the direction $\partial/\partial z$.
When we impose one null symmetry along $\partial /{\partial \bar z}$
and another along $\partial /{\partial \bar \lambda}$ we obtain
the Lax pair:
\[
L_0 = \frac{\partial}{\partial z} + A(z,\lambda),
\qquad  L_1 = \frac{\partial}{\partial \lambda} + B(z,\lambda).
\]

\noindent
{\bf Def\/inition 3.13.}
{\it Let $g = g_1 g_2$, then the Polyakov-Wiegmann formula is defined
by
\[
S_{WZW}(g_1 g_2) = S_{WZW}(g_1) + S_{WZW}(g_2) + \frac{1}{2\pi}\int
d^2z \langle g_{1}^{-1}{\bar \partial}g_1 , {\partial g_2}g_{1}^{-1}
\rangle .
\]}

Hence from our previous computation we can assert:

\medskip

\noindent
{\bf Proposition 3.14.}
{\it For $g = g_+g_{-}^{-1}$, the action becomes
\[
S = S^{u}(g_+) + S^{\bar v}(g_-) + \int d^2z
\langle A^u (g_+),A^{\bar v}(g_-)\rangle,
\]
where
\[
\ba{l}
\ds S^u (g_+) = 2\int d^2z \langle
u,{\bar \partial}g_+ g_{+}^{-1}\rangle  + S_{WZWN}(g_+),
\vspace{3mm}\\
\ds S^{\bar v}(g_-)  = 2\int d^2z
\langle v,{\partial}g_- g_{-}^{-1}\rangle + S_{WZWN}(g_{-}^{-1}).
\ea
\]
}

It is easy to see that $S = S(g_+ g_{-}^{-1})$ is a modif\/ied
Polyakov-Weigmann formula.
This action is gauge invariant under
\[
 g_+ \longrightarrow g_{+}h,  \qquad   g_- \longrightarrow g_- h,
\]
where $h$ is independent of $\lambda$ and the equation of motion is
a zero curvature equation with a spectral parameter on an arbitrary
Lie algebra ${\cal G}$: 
\[
{\bar \partial}A^u - {\partial A^{\bar v}} + [ A^u , A^{\bar v} ] = 0.
\]

This equation does not guarantee integrability
of the system. In particular, if we choose
\[
\ba{l}
A^{u} = A_0 + A_1 {\lambda} + A_2 {\lambda}^2 + \cdots + A_{n}{\lambda}^n,
\vspace{2mm}\\
\ds  A^{\bar v} = A^{u}{\lambda}^{-1},
\ea
\]
then one recovers the Adler-Kostant-Symes equation,
where $ A^{u}$ is considered to be a Lax operator $L$,
and $A^{\bar v}$
is the gradient of $H = \frac{1}{2}\mbox{tr}(L^2 {\lambda}^{-1})$.
In fact, the hierarchy of AKS system can be recasted into this
zero curvature form. 

\section{Applications}

In this section we will present some examples. We have already
stated that our Euler-Lagrange
equation is a zero curvature equation, and hence does not have an
immediate mechanical meaning. We show that after imposing the
Cartan decomposition of Lie algebras, we obtain a family
nontrivial mechanical systems associated to an arbitrary 
Riemannian symmetric space. 

\subsection{Periodic Toda Lattice }

Let $(\alpha_0, \alpha_1, \ldots ,\alpha_n )$ be a system of simple roots of
the af\/f\/ine Lie algebra ${\hat {\cal G}}$,
where $(\alpha_1, \ldots ,$ $\alpha_n )$
are simple roots of the original f\/inite dimensional algebra ${\cal G}$, and
$ -\alpha_0 = \sum\limits_{j=1}^{n} a_j \alpha_j $ is the highest root.

Let $(s_0, s_1, \ldots , s_n )$ be a set of non-negative integers without
a common divisor, and $ N = \sum\limits_{j=1}^{n} a_j s_j$
be the order of $\sigma$, where $\sigma^N = 1$.

\medskip

\noindent
{\bf Def\/inition 4.1.}
{\it A grading is a decomposition
$ {\cal G} = \oplus_{j \in Z} {\cal G}_j$ of the Lie algebra
${\cal G}$ into a direct sum of subspaces ${\cal G}_j$, such that
\[
\ba{l}
[ {\cal G}_i , {\cal G}_j ] \subset {\cal G}_{i+j} \quad \mbox{mod}\; N,
\vspace{2mm}\\
\sigma {\cal G}_k = {\epsilon}^k {\cal G}_k, \qquad
\epsilon = e^{\frac{2\pi i}{N}}.
\ea
\]
 This automorphism is called Coxeter automorphism.}

\medskip

An invariant subalgebra is a direct sum
\[
{\cal G}_0 = {\bf R} \oplus \cdots \oplus {\bf R} \oplus {\cal G}(k),
\]
where ${\cal G}(k)$ is a semi-simple subalgebra generated by simple roots
$( \alpha_{j_1}, \ldots , \alpha_{j_k} )$ for which
\[
 s_{j_1} = \cdots = s_{j_k} = 0.
\]

\noindent
{\bf Def\/inition 4.2.}
{\it When $ s_0 = s_1 = \cdots = s_n = 1$,
${\cal G}_0 \cong {\cal H}$ is a Cartan
 subalgebra and $ N = \sum\limits_{j=0}^{n} a_j = h$ is the Coxeter number.}

\medskip

Let $( H_j , E_j , F_j)$  $\forall \; j = 0, \ldots, n$ be the Cartan Weyl
basis of ${\hat {\cal G}}$. We def\/ine the following action of $\sigma$:
\[
 \sigma E_j = {\epsilon}^{s_j} E_j, \qquad
  \sigma H_j = H_j, \qquad
\sigma F_j = {\epsilon}^{-s_j}F_j .
\]
Let us consider the zero curvature equation again for
\[
A^u = A_0 + A_1 \lambda, \qquad
 A^{\bar v} = A_{-1}{\lambda}^{-1}.
\]
Then the zero curvature equation decomposes into   
\[
\ba{ll}
(1) &     {\bar \partial}A_1 = 0,
\vspace{2mm}\\
(2) &  \partial A_{-1} = [ A_0 , A_{-1} ],
\vspace{2mm}\\
(3) & {\bar \partial}A_0 + [ A_1 , A_{-1} ] = 0.
\ea
\]
One can readily identify
\[
A_0 \in {\cal G}_0, \qquad  A_1 \in {\cal G}_1, \qquad
 A_{-1} \in {\cal G}_{N-1}.
\]
Moreover $ A_1 \in {\cal G}_1 $ is a constant matrix in ${\cal G}_1$.

\medskip

\noindent
{\bf Proposition 4.3.}
{\it Let $ \eta \in {\cal G}_{N-1}$ and $ A_0 = q^{-1}{\partial}q$.
Then the above system of equations reduces to the periodic Toda lattice
equation.}

\medskip

\noindent  
Sketch of the {\bf Proof:}
From the first equation we let $A_1 = \eta$ be a constant matrix
in ${\cal G}_1$ and, if $ \xi \in {\cal G}_{N-1}$,
then from the second equation we obtain $A_{-1} = q^{-1}\xi q$
 and $ A_0 = -q^{-1}{\partial q}$.
 Finally from the third equation we obtain
\[
{\bar \partial}q^{-1}{\partial}q -  [ \eta, q^{-1}\xi^{-1}q ] = 0,
\]
by the substitution of $ q = \exp(\phi)$, where $\phi \in {\cal H}$
yields the periodic Toda lattice equation.

\hfill $\blacksquare$

\subsection{Hermitian Symmetric Spaces and Integrablity}

A Riemannian manifold $M$ is called a globally symmetric Riemannian space,
if every point $ p \in M$ is a f\/ixed point of involutative isometry of $M$
which takes any geodesic through $p$ into itself as a curve but reverses its
paramatrization.

Let $G$ be a semi-simple Lie group and ${\cal G}$ its Lie algebra.
Let $M$ be a homogeneous space of $G$ such that $M$ is a dif\/ferentiable
manifold on which $G$ acts transitively. There is a homeomorphism
of the coset space $G/H$ onto $M$ for some isotropy subgroup $H$ at
a point of $M$. Let ${\cal H}$ be the Lie algebra of $H$ and ${\cal G}$
satisf\/y
\[
{\cal G} = {\cal H} \oplus {\cal M}  \qquad \mbox{and} \qquad
[ {\cal H}, {\cal H} ] \subset {\cal H},
\]
where ${\cal M}$ is a vector space complement of ${\cal H}$.
 Furthermore, if ${\cal H}$ and ${\cal M}$ satisfy
$ [ {\cal H},{\cal M} ] \subset {\cal M} $ then $G/H$ is called
the reductive homogeneous space.
We can associate to these spaces a canonical connection with curvature and 
torsion. Curvature and torsion at a f\/ixed point $p \in G/H$ are
given purely in terms of the Lie bracket:
\[
\ba{l}
 (R(X,Y)Z)_p = -[[X,Y]_{\cal H},Z] \qquad \forall \; X,Y,Z \in {\cal M},
\vspace{2mm}\\
T(X,Y)_{p} = -[X,Y]_{\cal M}  \qquad \forall \;  X,Y \in {\cal M}.
\ea
\]

\noindent
{\bf Def\/inition 4.4.}
{\it A Hermitian symmetric space is a coset space $G/H$ for Lie groups whose
associated Lie algebras are ${\cal G}$ and ${\cal H}$ , with the decomposition
\[
{\cal G} = {\cal H} \oplus {\cal M}
\]
which satisfy the commutation relations
\[
 [ {\cal H}, {\cal H} ] \subset {\cal H},
 \qquad  [ {\cal H}, {\cal M} ] \subset {\cal M},
 \qquad  [ {\cal M}, {\cal M} ] \subset {\cal H}.
\] }

For Hermitian symmetric spaces the curvature satisf\/ies
\[
(R(X,Y),Z)_p = -[[X,Y],Z] \qquad \forall \;
X,Y,Z \in {\cal M},
\]
here $[X,Y] \in {\cal H}$ is satisf\/ied automatically due to
$[{\cal M},{\cal M}]$.

Let $k$ be an element in the Cartan subalgebra of ${\cal G}$, whose
centralizer in ${\cal G}$ is
\[
{\cal H} = \{ l \in {\cal G}: [k,l] = 0 \}.
\]
Let $ j = \mbox{ad}\; k = [ k , \ast ]$ be a linear map
\[
j : T^{\ast}(G/H) \longrightarrow T^{\ast}(G/H)
\]
satisfying $ j^2 = -1$ or $ [k,[k,m]] = -m $ for $m \in M$.

Let us consider again the zero curvature equation
\[
{\bar \partial}A^u - {\partial A^{\bar v}} + [ A^{u} , A^{\bar v} ] = 0.
\]

At this stage we project the zero curvature equation into
the real plane and treat ${\bar \partial} = {\partial}/{\partial x}$
and $\partial = {\partial}/{\partial t}$.
We assume that $A^{u}$ is the orbit $L$ through $u = \lambda^3 A$,
where $A = i \;\mbox{diag}\; (1,-1)$ is an
Cartan element of $su(2)$.
 Let us derive the orbit via the coadjoint action
\[
L = B^{-1}(\lambda^3 A) B,
\]
where
\[
B = \prod_{i=1}^{4} (b_i \lambda^{-i}, e^{\beta_i}\lambda^{-i}).
\]
Here $b_i$'s are central elements and $\beta_i \in {\cal M} $.
After an elaborated computation we obtain
\[
 L = \lambda^3 A + \lambda^2 Q + \lambda
 \left( P - \frac{i}{2}[Q_- , Q_+]\right) + T + [ S,Q ],
 \]
where
\[
\ba{l}
Q = [ A , \beta_1 ],
\vspace{2mm}\\
\ds P = [ A , \beta_2 ] + \frac{1}{2} [ Q , \beta_1 ],
\vspace{2mm}\\
\ds  T = [ A , \beta_3 ] + \frac{1}{2} [[ A , \beta_1], \beta_2 ]
+ \frac{1}{2}[[ A , \beta_2 ], \beta_1 ] + 
\frac{1}{6} [[ Q , \beta_1], \beta_1 ],
\vspace{2mm}\\
\ds S = \frac{i}{2} [ P_+ - P_- ] + cQ.
\ea
\]
If we assume $ H = -\frac{1}{8} \mbox{tr} ( L^2 \lambda^{-2})$,
then
\[
A^{\bar v} ( = \pi_+ \; \mbox{grad}\; H) = -\pi_+
\frac{1}{4}L \lambda^{-2} = -\frac{1}{4}(A\lambda + Q).
\]
  
\noindent
{\bf Proposition 4.5.}
{\it Let $({\cal O}_u , \omega_u)$ be the symplectic orbit,
where $\omega_u$ is the Killing two form on the orbit.
Then the Hamiltonian equations of motion, corresponding to 
$ H(L) = -\frac{1}{8}\mbox{\rm tr} (L^2 \lambda^{-2})$,
generates the system
of third order partial differential equations in ${\bf R}^n$.}

\medskip

 In this case the zero curvature equation is
\[
\frac{dL}{dt} = [ A\lambda + Q , L ] + \frac{1}{4}(A\lambda + Q)_x.
\]
Setting various coef\/f\/icients of $\lambda^m$ equal to zero we obtain:
\[
{\dot Q} = [ A, P ] - \frac{i}{2} [ A , [ Q_- , Q_+ ]].
\]
We now apply the group decomposition properties of Hermitian
symmetric spaces. Observe that $[ Q_- , Q_+ ] \in h$ and that 
$A$ is a constant matrix. Hence we obtain
\[
P = -\frac{i}{2} ({\dot Q_+} - {\dot Q_-}).
\]
Similarly
\[
\ba{l}
\ds  T = -\frac{1}{4}{\ddot Q} + \frac{1}{4} [ Q_+ , [ Q_- , Q_+]]
- \frac{1}{4} [ Q_-, [Q_- , Q_+]],
\vspace{3mm}\\
\ds  S = \frac{1}{4} ({\dot Q_+} + {\dot Q_-}) + cQ.
\ea
\]
  
Finally equating the $\lambda^0$ coef\/f\/icient we obtain
\[
 {\dot T} + [ S , Q ]_t = [ Q , T ] + [ Q , [ S , Q]] + \frac{1}{4}Q_x.
\]
If we choose 
\[
 Q = \left (\matrix{0 & q^{\dag}\cr
-q & 0\cr}\right),
\]
then we get, from the zero curvature equation,
\[
q_{ttt} + 6q_t |q|^2 + q_x = 0.
\]
This is a coupled KdV equation.

 When we consider the orbit $L$ through $ u = \lambda^2 A$
 we obtain
\[
 L = \lambda^2 A + \lambda Q + \left( P - \frac{i}{2}[ Q_- , Q_+ ]\right).
\]
A similar calculation yields the nonlinear Schr\"odinger
equation
\[
q_{tt} + iq_x + 2q|q|^2 = 0.
\]

\subsection{Geometric Action and Virasoro Group}

The geometric action of the Virasoro group has the form
of Polyakov's 2d quantum gravity
\[
S_{\mbox{\scriptsize grav}} = \int d^2z \frac{{\bar \partial}F}{\partial F}
\left(\frac{{\partial}^3F}{\partial F} - 2\left(\frac{{\partial}^2F}
{\partial F}\right)^2\right),
\qquad  (F \in \; \mbox{Dif\/f}\; (S^1)).
\]
Let $S^1$ be the circle parametrized by $x : 0 \leq x \leq 2\pi $
and $\mbox{Dif\/f}\; (S^1)$ be the group of all orientation preserving
$C^{\infty}$ dif\/feomorphisms of $S^1$. It is natural to consider
the Lie algebra of vector f\/ields on $S^1$ $\mbox{Vect}\; (S^1)$ as
its Lie algebra.
The dual of the $\mbox{Dif\/f}(S^1)/S^1$
is identif\/ied with the space of quadratic
dif\/ferential forms $u(x)dx^2$ by the following pairing
\[
\langle  u(x) , \xi \rangle = \int_{0}^{2\pi} u(x)\xi (x) dx,
\]
 where $ \xi = \xi (x)\frac{d}{dx} \in  \mbox{Vect}\; (S^1).$

Let us consider the shift of $(u(x),c)$, induced by $S^1$
dif\/feomorphism
\[
\ba{l}
\ds x \longrightarrow s(x) = x + \epsilon f(x),
\vspace{2mm}\\
\ds  (u(x),c) \longmapsto {s^{\prime}(x)}^{3/2}
(u(s(x)),c){s^{\prime}(x)}^{1/2} = ({\tilde u(x)}, c),
\ea
\]
where
\[
{\tilde u(x)} = {s^{\prime}(x)}^2 u(s(x)) + \frac{1}{2}
 \left(\frac{s^{\prime \prime \prime}}{s^{\prime}} -
\frac{3}{2}\left(\frac{s^{\prime \prime}}{s^{\prime}}\right)^2\right).
\]
The last term is known as Schwarzian ${\cal S}(s)$. After redef\/ining, or
adjusting, the coef\/f\/icients we can def\/ine the
current
\[
A = u(F)(\partial F)^2 - \frac{c}{24\pi}{\cal S}(F).
\]

In this case we have the action in the form of $ S = S^u (A) - H(A)$,
where 
\[
 S^u = -\int d^2 z u(F) {\partial F}{\bar \partial} F  +
 \frac{c}{48\pi} S_{\mbox{\scriptsize grav}},
\]
which is linear with respect to $A$.

The equation of motion is
\[
{\bar \partial}\; \mbox{Ad}_{F}^{\ast}(u,c) = \mbox{ad}_{v}^{\ast}\;
\mbox{Ad}_{F}^{\ast}(u,c),
\]
where $ v = \mbox{grad}\; H \in  \mbox{Vect}\; (S^1).$

This action can be derived from the canonical action on the cotangent bundle
of the group $\hat {\rm D}\mbox{if\/f}\;  S^1$.
The above equation can be transformed
 to symmetric form by the Polyakov-Wiegmann formula for the group
 $ \mbox{Dif\/f} \; S^1$.
 Unfortunately this equation can not be recasted to the zero curvature
equation.
Nevertheless, some integrable models can be described within
this approach.
 
\section{Appendix}
In this appendix we present a connection between
 the nonlinear Schr\"odinger equation and 
 the continuous Heisenberg ferromagnetic equation.
The continuous Heisenberg ferromagnetic model is an important integrable
model associated to some Hermitian symmetric spaces~[17, 18].

The action of the Heisenberg ferromagnetic model is given by
\[
S = \int d^2z \; \mbox{tr} \left[ 2u{\bar \partial}gg^{-1} +
\partial (g^{-1}kg) \partial (g^{-1}kg)\right],
\]
where $ k \in {\cal H}$ is a constant element.

Let us def\/ine $ Q := g^{-1}kg$, which  immediately leads to

\medskip

\noindent
{\bf Lemma 5.1.}
\[
{\partial Q} = [ Q , g^{-1}\partial g ],
\qquad 
{\bar \partial}Q = [ Q , g^{-1}{\bar \partial}g ].
\]

 The equation of motion is
\[
 {\bar \partial}Q + \partial [ Q , \partial Q ] = 0.
\]

\noindent
{\bf Lemma 5.2.}
{\it When $ Q = g^{-1}kg \in {\cal M}$, then ${\bar \partial}Q +
\partial [ Q , {\partial Q}] = 0$ is gauge equivalent to
\[
 \partial ({\partial g}g^{-1}) - [ k, {\bar \partial}gg^{-1} ] = 0.
\]}

\noindent
Sketch of the {\bf Proof:}
\[
 [ Q , {\partial Q}] = [ Q , [ Q , g^{-1}\partial g]]
 = g^{-1} [ k , [ k , {\partial g}g^{-1}]g
 = - g^{-1}({\partial g}g^{-1})g = -g^{-1}{\partial g}.
\]

It is easy to prove
\[
 \partial (g^{-1}\partial g ) = g^{-1}\partial ({\partial g}g^{-1})g
\].
 Then the result follows from these two.
 \hfill $\blacksquare$

\medskip

Additionally we have an identity 
\[
[ {\bar \partial} + {{\bar \partial}g}g^{-1} ,
\partial + {\partial g}g^{-1}] = 0.
\]
There is a natural splitting
\[
{\bar \partial}g g^{-1} = ({{\bar \partial}g}g^{-1})_{\cal M}
+ ({{\bar \partial}g}g^{-1})_{\cal H},
\]
where the subscripts ${\cal M}$ and ${\cal H}$ refer
to the component of ${{\bar \partial}g}g^{-1}$ in these vector
subspaces.

Hence the above equation reduces to
\[
[ k , ({{\bar \partial}g}g^{-1})_{\cal M} ] - \partial ({\partial g}g^{-1})
= 0.
\]

\medskip

\noindent
{\bf Lemma 5.3.}
\[
[ k , \partial ({\partial g}g^{-1}) ] = -({{\bar \partial}g}g^{-1})_{\cal M}.
\]

\medskip

\noindent
Sketch of the {\bf Proof.} We know
\[
\ba{l}
[ k , ({{\bar \partial}g}g^{-1})_{\cal M} ] =
\partial ({\partial g}g^{-1}),
\vspace{2mm}\\
{}[ k , [ k , ({{\bar \partial}g}g^{-1})_{\cal M} ]]
 = [ k , \partial ({\partial g}g^{-1}) ].
\ea
\]

\medskip

\noindent
{\bf Lemma 5.4.}
\[
[ k , \partial ({\partial g}g^{-1}) ] = -({{\bar \partial}g}g^{-1})_{\cal M}.
\]

\medskip

\noindent
Sketch of the {\bf Proof.} We know
\[
\ba{l}
[ k , ({{\bar \partial}g}g^{-1})_{\cal M} ] =
 \partial ({\partial g}g^{-1}),
 \vspace{2mm}\\
{}[ k , [ k , ({{\bar \partial}g}g^{-1})_{\cal M} ]]
 = [ k , \partial ({\partial g}g^{-1}) ],
\ea
\]
where we have applied $\mbox{ad} \; k$ on both sides.
\hfill $\blacksquare$

\medskip

Let us decompose the zero curvature equation into ${\cal H}$
and ${\cal M}$ part:
\[
\partial ({{\bar \partial}g}g^{-1})_{\cal H} + [ {\partial g}g^{-1} ,
({{\bar \partial} g}g^{-1})_{\cal M} ] = 0,
\]
and
\[
\partial ({{\bar \partial}g}g^{-1})_{\cal M} - 
{{\bar \partial}g}({\partial g}g^{-1}) + [ {\partial g}g^{-1} ,
({{\bar \partial}g}g^{-1})_{\cal H} ] = 0
\]
respectively.

From the ${\cal H}$ part of the equation we obtain
\[
\ba{l}
\ds \partial ({{\bar \partial}g}g^{-1})_{\cal H} - [{\partial g}g^{-1} ,
[ k , \partial ({\partial g}g^{-1}) ] = 0,
\vspace{2mm}\\
\ds \partial ({{\bar \partial}g}g^{-1})_{\cal H} + \frac{1}{2}
\partial [ {\partial g}g^{-1} , [ {\partial g}g^{-1} , k ]] = 0,
\ea
\]

Hence we obtain $({{\bar \partial}g}g^{-1})_{\cal H}$ upto
some constant which we can always set to zero:
\[
({{\bar \partial}g}g^{-1})_{\cal H} = -\frac{1}{2} [ {\partial g}g^{-1},
[{\partial g}g^{-1} , k ]].
\]

We f\/inally derive the nonlinear Schr\"odinger equation
\[
{\bar \partial}({\partial g}g^{-1}) + [ k , \partial^2 ({\partial g}g^{-1})]
+ \frac{1}{2} [ {\partial g}g^{-1} , [ {\partial g}g^{-1} , [ {\partial g}
g^{-1}, k ] ]] = 0.
\]

\subsection*{Acknowledgement}
We thank the Max Planck Institut f\"ur Mathematik,
Bonn, for their kind hospitality and providing an excellent working 
condition during the initial stages of this
work. One of us (PG) is also grateful to the organisers of the
``Non-Perturbative Aspects of Quantum Field Theory'' held at
Isaac Newton Institute, Cambridge, and I.H.E.S.
for their hospitality during the later stages of this work.

\label{guha-lp}

\end{document}